# Merits of Heavy-Heavy Molecules for Electron Electric Dipole Moment Searches

A. Sunaga*, M. Abe, and M. Hada
*Tokyo Metropolitan University, 1-1, Minami-Osawa, Hachioji-city, Tokyo 192-0397, Japan*

B. P. Das
*Department of Physics and International Education and Research Center of Science
Tokyo Institute of Technology, 2-12-1-H86 Ookayama, Meguro-ku, Tokyo152-8550, Japan*

The electric dipole moment of the electron (eEDM) and the Scalar-PseudoScalar (S-PS) interaction are probes of new physics beyond the standard model of elementary particles, but experiments to observe them using atoms and molecules are still in progress. Molecules that have a large effective electric field ($E_{\text{eff}}$), S-PS coefficient ($W_s$), and permanent electric dipole moment (PDM) are in principle favorable candidates for such experiments, and hence, it is necessary to analyze these properties. In this work, we calculate $E_{\text{eff}}$, $W_s$, and PDM for Ra systems; RaF, RaX (X = Cl, Br, I, and At) and RaY (Y = Cu, Ag, and Au) using the Dirac-Fock and the relativistic coupled-cluster methods. We find that RaX and RaY have larger $E_{\text{eff}}$ and $W_{s,\text{Ra}}$ than RaF. We explain this finding by taking into consideration the large *s-p* mixing for RaX and RaY, similar to what we had done in our previous work using hydrides and fluorides (A. Sunaga *et al.*, Phys. Rev. A **95**, 012502 (2017)). We also discuss the suitability of RaX and RaY molecules for eEDM experiments from the viewpoint of their large PDM and small polarizing electric field ($E_{\text{pol}}$).

## I. INTRODUCTION

An electric dipole moment of an electron (eEDM) arises from the violations of parity (*P*) and time (*T*) reversal symmetries. The value of the eEDM ($d_e$) in the standard model (SM) is very small, the order of 10 to −38 *e* cm [1–3], but many of the particle physics models beyond the SM (BSM) predict values that are many orders of magnitude larger [1–3]. The observation of a non-zero eEDM would be an unambiguous signature of the BSM.

The Scalar-pseudoscalar (S-PS) interaction between nucleons and electrons is a *P*, *T*-odd interaction like the eEDM interaction, and its coupling constant is referred to as the S-PS interaction constant ($k_s$) [4,5]. The S-PS interaction is also a probe of new physics beyond the SM, and it has been studied in extensions of the SM, like the multi-Higgs models; e.g. the minimal supersymmetric standard model [6,7] and the aligned two-Higgs-doublet model (A2HDM) [8]. The energy shift due to the eEDM and the S-PS interactions could be observed in an experiment using a paramagnetic system.

The two roles of the theoretical calculation for the eEDM search are i) determining the effective electric field ($E_{\text{eff}}$) and the S-PS coefficient ($W_s$) accurately for some target molecules, and ii) proposing candidate molecules suitable for eEDM searches. The focus of this paper are related to the latter one.

Since the sensitivity of the experiment increases as $E_{eff}$ and $W_s$ become large, it is important to understand the mechanisms that enhance $E_{eff}$ and $W_s$. It appeared from the early work of Sandars in 1964 [9] that molecules with large electric polarization have larger effective electric fields. This idea is now commonly accepted [10–12]. However, recently, we reported that YbH and HgH have larger $E_{eff}$ than YbF and HgF, although their polarizations are smaller [13]. In our previous work, diatomic molecules are not understood as "heavy atoms in the electric field, which is produced by the electron moving to the lighter atom". The mechanism of s-p mixing is analyzed based on the orbital interaction theory. Our idea that non-fluorides with weaker polarization also have large $E_{eff}$ can be extended to different kinds of molecules. Gaul et al. calculated $E_{eff}$ and $W_s$ for several hydrides, nitrides, oxides, and fluorides, and showed that non-fluorides could also have large $E_{eff}$ and $W_s$ [14].

In view of the above background, diatomic molecules containing two heavy atoms (heavy-heavy molecules) become attractive. Even though the electronegativities of heavy atoms are much smaller than that of fluorine atom, heavy-heavy molecules might have large $E_{eff}$. If the effective electric fields coming from two heavy atoms constructively contribute to the total molecular $E_{eff}$, then heavy-heavy molecules might have larger $E_{eff}$ than heavy-fluorides. In previous works, however, it has not been reported that heavy-heavy molecules have larger $E_{eff}$ than heavy-light systems. Meyer et al. proposes Alkali-Yb and Alkaline earth-Yb systems [11], but the signs of $E_{eff}$ coming from the two heavy atoms in the molecules are always opposite. As a result, the overall $E_{eff}$ decreases; e.g., $E_{eff}$(Yb) and $E_{eff}$(Sr) in YbSr$^+$ are -21.9 and 10.6 (GV/cm), respectively, and the overall $E_{eff}$ is -11.3 (GV/cm). Prasannaa et al. proposed HgBr and HgI [15], which are also heavy-heavy systems, but their $E_{eff}$ are smaller than HgF ($E_{eff}$ for HgF, HgBr, and HgI are 115.4, 109.3 and 109.3 (GV/cm) at the CCSD level, respectively).

The permanent dipole moment (PDM) of the molecule is another important quantity for the eEDM experiments using $^2\Sigma_{1/2}$ molecules. PDM is directly related to the orientation of the molecules, and a large PDM decreases the systematic error of the experiment. Many candidate molecules with $^2\Sigma_{1/2}$ includes halogens; e.g. BaF [16], YbF [17], HgX (X = F, Cl, Br, and I) [15], PbF [18,19], RaF [20,21].

One of the reasons for this would be that molecules containing a halogen (especially, fluorine) have larger PDMs than other molecules, because of the large electronegativity of the halogen. However, PDM also depends on how much the polarized charges are separated in the molecule; i.e., PDM may depend on the molecular bond length. From this, heavy-heavy molecules, which have larger bond lengths than fluorides, may have larger PDM than fluorides. However, the contribution of the molecular bond length to PDM has never been discussed in the context of eEDM searches.

In this work, we have calculated $E_{eff}$, $W_s$, and PDM for RaF, RaX (X = Cl, Br, I, and At) and RaY (Y = Cu, Ag, and Au) molecules at the Dirac-Fock (DF) and the relativistic coupled cluster singles and doubles (RCCSD) methods. We find that all the RaX and RaY except for RaAu have larger $E_{eff}$ than RaF at the CCSD level. $E_{eff}$ of RaAu (55.6 GV/cm) is almost same as that of RaF (56.9 GV/cm). The value of $W_{s,Ra}$ for RaF is the smallest in our target molecules. The reason for this can be explained by the relatively smaller *s-p* mixing for RaF molecule. The characteristics of the *s-p* mixing is explained by the orbital interaction theory, as done in our previous work [13]. In addition to the analysis of $E_{\text{eff}}$ and $W_s$, we find that heavy-heavy systems can have larger PDM than fluorides. For example, RaAu has larger PDM (5.2 D) than RaF (4.5 D) at the CCSD level, despite the lower electronegativity of Au. The reason for the larger PDM would be the longer bond length of the heavy-heavy systems. Finally, we mention the suitability of heavy-heavy systems in the viewpoint of the polarizing electric field ($E_{\text{pol}}$), which is the applied external electric field for orienting the molecules in the experiment. We discuss potential of heavy-heavy molecules for the eEDM experiment from the viewpoints of the enhancement factors ($E_{\text{eff}}$ and $W_s$), and the orientation (PDM and $E_{\text{pol}}$), using Ra systems as examples.

## II. THEORY

The operator of the eEDM interaction is written as follows [22],

$$\hat{H}_{\text{eEDM}} = -d_e \sum_j^{N_e} \beta \boldsymbol{\Sigma}_j \cdot \mathbf{E}_{\text{int}}. \tag{1}$$

Here, $d_e$ is the value of the eEDM, $j$ is the label indices for electrons, $N_e$ is the number of the electrons in the molecule, $\beta$ is the Dirac matrix, $\boldsymbol{\Sigma}$ is the four-component Pauli matrix, and $\mathbf{E}_{\text{int}}$ is the total internal electric field, which is created by the nuclei and the electrons in the molecule.

The effective electric field ($E_{\text{eff}}$) is defined by the below equation.

$$E_{\text{eff}} = -\left\langle \Psi \left| \frac{\hat{H}_{\text{eEDM}}}{d_e} \right| \Psi \right\rangle, \tag{2}$$

where $\Psi$ is the four-component electronic wavefunction of the molecule. In this work, we used the below one-electron operator for the calculation of $E_{\text{eff}}$ [23,24].

$$E_{\text{eff}} = -2ic \left\langle \Psi \left| \sum_j^{N_e} \beta \gamma_5 \mathbf{p}_j^2 \right| \Psi \right\rangle, \tag{3}$$

where $i$ is the imaginary unit, $c$ is the speed of light, $\gamma_5$ is the Dirac matrix, and **p** is the momentum operator.

The S-PS interaction is defined by the following operator [4,5]:

$$\begin{aligned}\hat{H}_{\text{S-PS}} &= \sum_A^{N_n} \hat{H}_{\text{S-PS},A} \\ &= \sum_A^{N_n} i \frac{G_F}{\sqrt{2}} k_{s,A} Z_A \sum_j^{N_e} \beta \gamma^5 \rho_A(\mathbf{r}_{Aj})\end{aligned}, \tag{4}$$

$N_n$ represents the total number of the nuclei, and $A$ is the label indices for the nuclei. $Z$ is the nuclear charge. $k_{s,A}$ is the dimensionless S-PS interaction constant of the atom $A$. We used the Gaussian-type distribution function for the normalized nuclear charge density $\rho$, as we had in our previous work [25]. The S-PS coefficient $W_{s,A}$ is defined for molecules with $^2\Sigma$ character as follows,

$$W_{s,A} = 2\left\langle \Psi \left| \frac{\hat{H}_{\text{S-PS},A}}{k_{s,A}} \right| \Psi \right\rangle. \tag{5}$$

The features common to $E_{\text{eff}}$ and $W_s$ are that they both depend on parity-odd interactions, and that the electronic wavefunction in the region close to the heavy nucleus mainly contributes to these properties. From these considerations, the mixing of the heavy atomic *s* and *p* orbitals (*s-p* mixing) increases the values of $E_{\text{eff}}$ and $W_s$. In addition, only the singly occupied molecular orbital (SOMO) contributes to the values of $E_{\text{eff}}$ and $W_s$ at the Kramers restricted Dirac-Fock level, because they are time-reversal odd properties. In the latter sections, we explain the mechanism for the enhancement of $E_{\text{eff}}$ and $W_s$ from the viewpoints of the *s-p* mixing in SOMO.

The permanent dipole moment of the molecule (PDM) is obtained using the below expression.

$$\text{PDM} = -\left\langle \Psi \left| \sum_i^{N_e} \mathbf{r}_i \right| \Psi \right\rangle + \sum_A^{N_n} Z_A \mathbf{R}_A, \tag{6}$$

where **r** and **R** are the position vectors of the electrons and nuclei.

We calculated the above molecular properties at the Dirac-Fock (DF) and the relativistic coupled cluster singles and doubles (RCCSD) level. The coupled-cluster wave function |ψ> is given by

$$|\psi\rangle = e^{\hat{T}}|\psi_0\rangle, \quad (7)$$

where |ψ₀> is the reference wavefunction, which is obtained at the DF level. At the RCCSD level, the cluster operator $\hat{T}$ is truncated as $\hat{T} \approx \hat{T}_1 + \hat{T}_2$. For the calculations of the expectation value of $\hat{O}$, we consider only the linear terms in the CCSD wave function as follows [26].

$$\langle\psi_0|(1+\hat{T}_1+\hat{T}_2)^{\dagger}\hat{O}_N(1+\hat{T}_1+\hat{T}_2)|\psi_0\rangle_C + O_0, \quad (8)$$

where $\hat{O}_N$ is the normal-order operator, the subscript $C$ refers to connected terms, and $O_0$ is the expectation value for the operator $\hat{O}$ at the DF level [27,28].

### III. COMPUTATIONAL METHOD

We use the Dirac-Coulomb Hamiltonian in all the calculations to obtain the molecular electronic wavefunction. We use the UTCHEM [29–31] and DIRAC08 [32], for the generation of DF orbitals and the molecular integral transformation, and the CCSD amplitudes, respectively. The two codes were modified by Abe *et al.* [24] and Sunaga *et al.* [25] for the calculations of the above molecular properties. The kinetic balance condition [29,33] is applied to the 2-spinor basis sets of the large and small components.

Dyall 2zp basis sets [34–38] were employed for the optimization of the bond length, while for the calculation of the molecular properties, we used Dyall ae2z basis sets [34–38] for all the elements. In the CCSD calculations, the cut off energy for the virtual orbitals was 80 a.u.

All the bond lengths for our target molecules were optimized by the spin-orbit exact two-component (X2C) method [39] and the density functional theory (DFT) except for RaAu molecule. For RaAu, we used the Dirac-Coulomb DFT with an approximation of two-electron integrals for small components [40], because there was a convergence problem using X2C. For the DFT calculations, PBE0 functional [41], and DIRAC17 [42] code is used. For the calculation of $W_s$, we choose the following isotopes; $^{223}$Ra, $^{19}$F, $^{35}$Cl, $^{79}$Br, $^{127}$I, $^{210}$At, $^{63}$Cu, $^{107}$Ag, and $^{197}$Au, and employ the experimental root-mean-squares (RMS) charge radii [43] except for At. We employ the RMS charge radii empirically obtained from the nuclear masses using Eq. (20) in ref. [44] for At, because the experimental RMS is not available.

### IV. NUMERICAL RESULTS

We show the values of the optimized bond lengths, $E_{\text{eff}}$, $W_s$, and PDM in Table I. (Note that $E_{\text{eff}}$ and $W_{s,\text{Ra}}$ are intentionally written as plus sign throughout this paper to discuss their magnitudes clearly. If we follow the same notation of the previous works for RaF [14,45], where $E_{\text{eff}}$ and $W_{s,\text{Ra}}$ are written as minus sign, all of the signs of $E_{\text{eff}}$ and $W_s$ shown in our paper will become opposite.) The correlation effect increases the absolute values of $E_{\text{eff}}$, $W_{s,\text{Ra}}$, $W_{s,\text{X}}$, and $W_{s,\text{Y}}$ for all of our molecules. The order of the values among our target molecules is the same at the DF and CCSD levels for each property, except for the $E_{\text{eff}}$ for RaF and RaAu. The absolute value of $E_{\text{eff}}$ (GV/cm) is larger in RaF than in RaAu at the DF level (43.4 and 50.9, respectively), while this becomes opposite (56.9 and 55.6) at the CCSD level.

Our values of RaF are in broad agreement with the absolute values reported by Sasmal *et al.* using the Dirac-Coulomb Z-vector CCSD method and Dyall cv4z basis set [21]. Their $|E_{\text{eff}}|$, $|W_s|$, and PDM are 52.5 GV/cm, 141.2 kHz and 3.8459 Debye, respectively. However, our bond length (2.30 Å) is slightly different from their bond length (2.24 Å). When we employ 2.24 Å for RaF, the values of $E_{\text{eff}}$, $W_{s,\text{Ra}}$, $W_{s,\text{F}}$, and PDM are 55.8 GV/cm, 149.6 kHz, 1.4 ×10⁻³ kHz, and 4.1 Debye at the CCSD level. The values of $E_{\text{eff}}$ and $W_s$ are almost the same as those in Table I. The value of PDM is clearly smaller than that in Table I (4.5 Debye). However, this tendency that the system with a small bond length (2.24 Å) has a small PDM is consistent with the discussion below in section V. From Table I, the values of $E_{\text{eff}}$ and $W_{s,\text{Ra}}$ for RaX and RaY are larger than RaF, except for $E_{\text{eff}}$ for RaAu at the CCSD level. In RaX molecules, the values of $E_{\text{eff}}$ and $W_{s,\text{Ra}}$ increase as the atomic number Z of X becomes larger. This tendency is opposite to that of the case for Hg halides [15]. In contrast to RaX, the values of $E_{\text{eff}}$ and $W_{s,\text{Ra}}$ for RaY decrease as the atomic number of Y becomes larger. RaAu has the smallest $E_{\text{eff}}$ at the CCSD level, while its $W_{s,\text{Ra}}$ is relatively large.

As for PDMs, all the RaX molecules have larger PDM than RaF at both the DF and CCSD levels. This tendency is opposite to that of the case of Hg halides [15]. In RaY, the PDMs for RaAg and RaAu at the CCSD level are larger than RaF. The PDMs for RaAg and RaAu at the DF level (2.8 and 3.6 Debye) are large enough for eEDM experiments, compared to the PDM of HgBr and HgI (2.71 and 2.06 Debye at the finite field CCSD level [46]). The correlation effects increase the PDMs for RaY much more.

In the below sections, we will discuss the following two points: (i) Why do heavy-heavy systems have larger $E_{\text{eff}}$ and $W_{s,\text{Ra}}$ than RaF? (ii) Why do heavy-heavy systems have larger PDM, although the electronegativities of X and Y atoms are smaller than F atom?

## V. ANALYSIS of $E_{\text{eff}}$ and PDM

We separate the contribution of $E_{\text{eff}}$ from each nucleus in the molecules at the DF level as follows.

$$E_{\text{eff},A} \approx -4ci\left( \sum_{k}^{N_{A,s}^L} \sum_{l}^{N_{A,p}^S} C_{s,k}^{*L} C_{p,l}^S \left\langle s_{1/2,k}^L \left| \mathbf{p}^2 \right| p_{1/2,l}^S \right\rangle \right.$$
$$\left. + \sum_{m}^{N_{A,p}^L} \sum_{n}^{N_{A,s}^S} C_{p,m}^{*L} C_{s,n}^S \left\langle p_{1/2,m}^L \left| \mathbf{p}^2 \right| s_{1/2,n}^S \right\rangle \right), \quad (9)$$

where, k, l, m, and n are the labels for basis set spinors of $s_{1/2}^L$, $p_{1/2}^S$, $p_{1/2}^L$ and $s_{1/2}^S$, respectively, and C is a molecular orbital coefficient of the SOMO. $N_{A,n}^M$ stands for the numbers of basis sets and for example, $N_{A,s}^L$ is the number of the large-component basis sets of s orbital for the atom A. $E_{\text{eff},\text{Ra}}$ is the contribution from Ra's $s_{1/2}$ and $p_{1/2}$ orbitals to the total $E_{\text{eff}}$. $E_{\text{eff},\text{X}}$ and $E_{\text{eff},\text{Y}}$ are the contribution from $s_{1/2}$ and $p_{1/2}$ orbitals of X and Y atoms, respectively. We neglect the contribution from d, f, and g orbitals to $E_{\text{eff}}$ in this analysis because their populations in SOMO are very small.

We show the values of $E_{\text{eff},\text{Ra}}$, $E_{\text{eff},\text{X}}$ and $E_{\text{eff},\text{Y}}$ for RaX and RaY molecules in Table II. The values of $E_{\text{eff},\text{Ra}} + E_{\text{eff},\text{X(Y)}}$ are very close to the values of $E_{\text{eff}}$ at the DF level, which are given in Table I. This implies that the contribution from the cross term between Ra and X (Y) atoms are very small. This is consistent with the nature of the locality of $E_{\text{eff}}$; i.e., the electronic wavefunction in the region close to the heavy nucleus mainly contributes to $E_{\text{eff}}$. The signs for $E_{\text{eff},\text{Ra}}$, $E_{\text{eff},\text{X}}$ and $E_{\text{eff},\text{Y}}$ are same as those of $W_{s,\text{X}}$ and $W_{s,\text{Y}}$, as shown in Table I. From these results, the signs of $E_{\text{eff},\text{X(Y)}}$ and $W_{s,\text{X(Y)}}$ would depend on the group that X and Y atoms belong to.

In RaF and RaX, $E_{eff,Ra}$ and $E_{eff,X}$ add up constructively, but in RaY, there is cancellation between $E_{eff,Ra}$ and $E_{eff,Y}$. Even in spite of the latter feature, the $E_{eff}$ values for RaY, are larger than that of RaF at the DF level, as shown in Table I. From Table II, the large values of $E_{eff,Ra}$ for RaY overcome the cancellation between the $E_{eff}$ of the individual atoms that make up this molecule and as a result, RaY has a larger total $E_{eff}$ than RaF.

The reason why the contributions from X and Y atoms are small is that the SOMO electrons are localized at the Ra atom. Table III shows the results of the Mulliken population (MP) analysis [47] for the SOMO. MP indicates the number of electrons, which belong to each atomic orbital in the molecule. The value of "Ra total" also includes the contribution from $d$, $f$ and $g$ orbitals, but we do not give each of the values because they are very small. From the values of "Ra total" in the table, the SOMO electrons are located at the Ra atoms for all of our target molecules.

Next, we will discuss the tendency of $E_{eff,Ra}$. From Table II, $E_{eff,Ra}$ increases as X becomes heavier in RaX, while $E_{eff,Ra}$ decreases as Y becomes heavier in RaY. This tendency is the same as that of $W_{s,Ra}$, as shown in Table I. The reason for this tendency can be explained by the magnitude of $s$-$p$ mixing, as shown in Table III. In RaX, the magnitudes of $s$-$p$ mixing (i.e., the values of Ra's $p$ population) increase as X becomes heavier. In contrast, the magnitudes of $s$-$p$ mixing decrease as Y becomes heavier in RaY. This tendency is consistent with those of $E_{eff,Ra}$ and $W_{s,Ra}$.

The characteristics of $s$-$p$ mixing can be understood on the basis of the orbital interaction theory [48,49], as done in our previous work [13]. From this theory, the large contribution of the virtual $p$ orbitals to SOMO is explained by the small energy differences of the valence orbitals ($\Delta\varepsilon$) and the large overlap integrals ($S$) of each atom in the molecules. (Note that the small energy difference would not always increase $E_{eff}$, as explained in our previous paper [13].) The $s$-$p$ mixing trend in Table III is consistent with the values of the overlap integrals and the energy differences. Here, the orbital energies are obtained from atomic DF calculations using the GRASP2K code [50]. The overlap integrals were obtained by using the contracted Dyall 4Z basis sets. The reason why RaF has a smaller $E_{eff}$ than those of RaX and RaY is due to its small $s$-$p$ mixing, which would originate from the large $\Delta\varepsilon$ and the small $S$.

Next, we analyze the reason why RaX and RaY can have larger PDM than RaF, although the electronegativities of X and Y atoms are smaller than F atom. Table IV shows three values, (i) the atomic charge of Ra, which is obtained from total MP of Ra, (ii) the optimized bond lengths, which are same as those in Table I, (iii) the products between the MP and the bond length, which corresponds to classical PDMs. From Table IV, the order of the classical PDMs is the same as that of the PDMs at the DF level in Table I, except for RaCl and RaBr. From this, our classical model would be valid for understanding the PDMs. We now consider the explanation for the PDM trends. The atomic charge of Ra for RaF is the largest among all of the RaX molecules, as expected due to the largest electronegativity of the F atom. However, the bond lengths for RaX and RaY are larger than RaF, and they also contribute to the PDM. The advantage due to the longer bond lengths of the former molecules can overcome the relatively small sizes of their electric polarization.

We note that the reasonably large electric polarization of RaX and RaY also contributes to large PDM. From the above considerations, it is clear that RaX have larger PDMs than RaF, and the PDMs of RaY are also not very small. Since our model is classical, the trends of the classical PDMs in Table IV do not completely agree with the trends at the CCSD level in Table I. However, the reason why RaX and RaY can have large PDMs can be qualitatively understood by their large bond lengths. The above discussion could be extended to other heavy-heavy systems with large bond lengths. Systems with large PDM are not only suitable for the beam experiment but also for the molecules embedded in a solid matrix of inert gas atoms, proposed by Vutha *et al*. [51].

**VI. Rotational constant and $E_{pol}$**

In this section, we discuss the advantage of heavy-heavy molecules with $^2\Sigma$ character from the viewpoint of its orientation.

The minimum external electric field required for orienting molecules refers to the polarizing electric filed ($E_{pol}$). $E_{pol}$ for $^2\Sigma$ molecules can be shown as follows [52].

$$E_{pol} = \frac{B(J+1)(J+2) - BJ(J+1)}{D}$$
$$= \frac{2B(J+1)}{D} \quad . \quad (10)$$

Here, $B$ is the rotational constant, $J$ is the quantum number of the molecular rotation, and $D$ is the value of PDM. In the case of a molecular state with $J = 0$, $E_{pol}$ becomes $2B/D$. Since a small external field is important to decrease the systematic error, molecules with small $E_{pol}$ are suitable for experiments. In other words, molecules with large $D$ and small $B$ are suitable for experiments in the case of $^2\Sigma$ molecules.

Table V shows the values of $E_{pol}$, calculated from the values of PDM at the CCSD level in Table I. As mentioned by Meyer *et al.* [11], heavy-heavy molecules have small $B$ due to larger reduced mass, and hence they have smaller $E_{pol}$. Our results are consistent with this point that they have made, but additionally, RaX and RaY have large PDMs as shown above, and hence they have much smaller $E_{pol}$ than RaF. The values of $E_{pol}$ of RaX and RaY (e.g. 0.34, 0.25, and 0.40 (kV/cm) for RaI, RaAt, and RaAu, respectively) are one- order magnitude smaller than those of the systems proposed previously; HgBr (~2 kV/cm), HgI (~2 kV/cm) [15], YbSr (5.5 kV/cm), YbRb (3.5 kV/cm) [11], and YbF (8.3 kV/cm) [53].

**VII. Conclusion**

In this paper, we find three results. i) RaX and RaY have larger $E_{eff}$ and $W_{s,Ra}$ than RaF. This can be explained on the basis of the larger *s-p* mixing in RaX and RaY. In the case of RaY, although the signs of $E_{eff,Ra}$ and $E_{eff,Y}$ are opposite, the cancellation between the two terms is not significant as the magnitude of $E_{eff,Y}$ is relatively small. ii) RaX and RaY have larger PDMs than that of RaF, which can be understood by the larger bond lengths of the former two molecules. iii) RaX and RaY have small $E_{pol}$, due to their small rotational constants and large PDMs.

We demonstrate that the orbital interaction theory can be a powerful tool even for heavy-heavy systems, to understand the tendency of the *s-p* mixing of SOMO. Also, heavy-heavy molecules can have larger PDM than fluorides when their relatively large bond lengths overcome their smaller electronic polarization. The production of sufficiently large numbers of RaX and RaY molecules for performing eEDM experiments with them would be challenging (The experiment using RaAg molecules is in preparation [54]).

However, our idea that heavy-heavy molecules have larger $E_{\text{eff}}$, $W_s$, and PDM than fluorides and their analysis, will be helpful for suggesting candidate molecules for eEDM experiments.

**ACKNOWLEDGMENTS**

We would like to thank Profs. V. Amar, K. Dyall, and T. Fukuyama for valuable discussion. This work was supported by JSPS KAKENHI Grant No. 17J02767, No. 17H03011, No. 17H02881, and No. 18K05040.

Table I. Summary of our calculation results ($E_{eff}$, $W_s$, $W_s/E_{eff}$, and PDM) at the DF and CCSD levels.

|  | Method | RaF | RaCl | RaBr | RaI | RaAt | RaCu | RaAg | RaAu |
|---|---|---|---|---|---|---|---|---|---|
| bond length (Å) | PBE0 | 2.30 | 2.82 | 2.97 | 3.23 | 3.33 | 3.01 | 3.23 | 3.00 |
| $E_{eff}$ (GV/cm) | DF | 43.4 | 47.6 | 48.8 | 51.5 | 56.7 | 56.2 | 55.3 | 50.9 |
|  | CCSD | 56.9 | 62.7 | 63.5 | 67.5 | 78.6 | 77.1 | 73.7 | 55.6 |
| $W_{s, Ra}$ (kHz) | DF | 116.9 | 127.9 | 130.9 | 137.3 | 142.3 | 150.9 | 149.3 | 146.8 |
|  | CCSD | 152.5 | 168.1 | 169.4 | 177.5 | 188.9 | 207.4 | 201.8 | 187.0 |
| $W_{s, X}$, $W_{s, Y}$ (kHz) | DF | $4.8 \times 10^{-4}$ | 0.01 | 0.1 | 0.6 | 9.7 | -0.03 | -0.3 | -7.9 |
|  | CCSD | $1.6 \times 10^{-3}$ | 0.02 | 0.2 | 1.5 | 20.7 | -0.2 | -1.6 | -30.6 |
| $W_{s, Ra}$(kHz)/$E_{eff}$(GV/cm) | DF | 2.7 | 2.7 | 2.7 | 2.7 | 2.5 | 2.7 | 2.7 | 2.9 |
|  | CCSD | 2.7 | 2.7 | 2.7 | 2.6 | 2.4 | 2.7 | 2.7 | 3.4 |
| PDM (Debye) | DF | 4.1 | 5.7 | 5.8 | 6.5 | 6.3 | 2.3 | 2.8 | 3.6 |
|  | CCSD | 4.5 | 6.1 | 6.3 | 6.9 | 6.7 | 3.6 | 5.1 | 5.2 |

Table II. The contribution from each Ra, X, and Y atom to the total $E_{eff}$ (GV/cm).

|  | RaF | RaCl | RaBr | RaI | RaAt | RaCu | RaAg | RaAu |
|---|---|---|---|---|---|---|---|---|
| $E_{eff, Ra}$ | 43.2 | 47.3 | 48.4 | 50.7 | 52.6 | 55.8 | 55.2 | 54.2 |
| $E_{eff, X}$, $E_{eff, Y}$ | $5.1 \times 10^{-4}$ | $7.3 \times 10^{-3}$ | 0.1 | 0.5 | 3.8 | -0.04 | -0.2 | -3.7 |
| $E_{eff, Ra} + E_{eff, X}$, $E_{eff, Y}$ | 43.2 | 47.3 | 48.5 | 51.3 | 56.4 | 55.7 | 54.9 | 50.5 |

Table III Mulliken population (MP) of SOMO electron, AO energy differences ($\Delta\varepsilon$) between the 7$s$ orbital of Ra atom and the valence orbital of F, X, and Y atoms for the eight molecules, and the overlap integrals ($S$) between the 7$s$ and the valence orbitals for the eight molecules. The energies of valence orbitals of F, Cl, Br, I, At, Cu, Ag, Au, and Ra (2$p_{3/2}$, 3$p_{3/2}$, 4$p_{3/2}$, 5$p_{3/2}$, 6$p_{3/2}$, 4$s$, 5$s$, 6$s$, and 7$s$) were evaluated from the ground state of the neutral atoms by GRASP2K.

|  | RaF | RaCl | RaBr | RaI | RaAt | RaCu | RaAg | RaAu |
|---|---|---|---|---|---|---|---|---|
| Ra ($s$) | 0.86 | 0.84 | 0.83 | 0.82 | 0.82 | 0.55 | 0.59 | 0.69 |
| Ra ($p$) | 0.11 | 0.12 | 0.12 | 0.14 | 0.14 | 0.30 | 0.27 | 0.23 |
| Ra total | 1.00 | 1.00 | 0.99 | 0.99 | 0.99 | 0.89 | 0.89 | 0.95 |
| X,Y ($s$) | $-4 \times 10^{-5}$ | $4 \times 10^{-4}$ | $5 \times 10^{-4}$ | $6 \times 10^{-4}$ | $7 \times 10^{-4}$ | 0.11 | 0.11 | 0.04 |
| X,Y ($p$) | $3 \times 10^{-4}$ | $4 \times 10^{-3}$ | 0.01 | 0.01 | 0.01 | $3 \times 10^{-3}$ | $-5 \times 10^{-3}$ | 0.01 |
| X,Y total | $3 \times 10^{-4}$ | $4 \times 10^{-3}$ | 0.01 | 0.01 | 0.01 | 0.11 | 0.11 | 0.05 |
| $\Delta\varepsilon$ (a.u.) | 0.57 | 0.34 | 0.28 | 0.22 | 0.17 | 0.08 | 0.07 | 0.13 |
| $S$ | -0.03 | -0.10 | -0.12 | -0.16 | -0.18 | No value[a] | 0.42 | 0.41 |

[a] The contracted Dyall basis set for Cu is not reported.

Table IV. Atomic charges of Ra obtained from the Mulliken population analysis, optimized bond lengths, and the classical PDMs for our target molecules.

|  | RaF | RaCl | RaBr | RaI | RaAt | RaCu | RaAg | RaAu |
|---|---|---|---|---|---|---|---|---|
| Atomic charge of Ra | 0.82 | 0.70 | 0.66 | 0.65 | 0.61 | 0.33 | 0.46 | 0.58 |
| Bond length (Å) | 2.30 | 2.82 | 2.97 | 3.23 | 3.33 | 3.01 | 3.23 | 3.00 |
| Classical PDM (Debye) | 9.1 | 9.5 | 9.5 | 10.0 | 9.7 | 4.8 | 7.1 | 8.4 |

Table V. Rotational constants ($B$) and polarizing electric fields ($E_{pol}$).

|  | RaF | RaCl | RaBr | RaI | RaAt | RaCu | RaAg | RaAu |
|---|---|---|---|---|---|---|---|---|
| $B$ (cm$^{-1}$) | 0.180 | 0.070 | 0.033 | 0.020 | 0.014 | 0.038 | 0.022 | 0.018 |
| $E_{pol}$ (kV/cm) | 4.77 | 1.36 | 0.62 | 0.34 | 0.25 | 1.23 | 0.52 | 0.40 |